\documentclass[twocolumn,showpacs,preprintnumbers,prb]{revtex4}
\usepackage{graphics}
\usepackage{graphicx}
\begin{document}
\title{Theoretical study of electron states in Au chains on NiAl(110)}
\author{Mats Persson}
\email{tfymp@fy.chalmers.se} \affiliation{ Department of Applied
Physics, Chalmers/G\"oteborgs University, S-41296, G\"oteborg,
Sweden}
\date{\today}

\begin{abstract}
We have carried out a density functional study of unoccupied,
resonance states in a single Au atom, dimers, a trimer and
infinite Au chains on the NiAl(110) surface. Two inequivalent
orientations of the ad-chains with substantially different
interatomic distances were considered. From the study of the
evolution of the electron states in an Au chain from being
isolated to adsorbed, we find that the resonance states derive
from the 6$s$ states of the Au atoms, which hybridize strongly
with the substrate states and develop a $p$-like polarization. The
calculated resonance states and LDOS images were analyzed in a
simple tight-binding, resonance model. This model clarifies (1)
the physics of direct and substrate-mediated adatom-adatom
interactions and (2) the physics behind the enhancements of the
LDOS at the ends of the adatom chains, and (3) the physical
meaning of the "particle-in-box" model used in the analysis of
observed resonance states. The calculated effective mass and band
bottom energy are in good agreement with experimental data
obtained from scanning tunnelling spectroscopy.
\end{abstract}
\pacs{73.20.Hb, 73.21.Hb, 68.37.Ef}
\maketitle

\section{Introduction \label{sec:intro}}

A most interesting development in nanoscience has been the
possibility to tailor and analyze the electronic properties of
single nanostructures of adatoms on metal surfaces, This
development has been made possible by the unique capabilities of
the scanning tunnelling microscope to image and manipulate single
adatoms and to probe their electronic structure by tunnelling
spectroscopy. A pioneering example is provided by the quantum
corral experiments, in which atomic manipulations were used to
assemble closed atomic structures, which in turn act as two
dimensional resonance cavities for surface state electrons on a
metal surface~\cite{CroLutEig93a,HelCroLutEig}. Another most
interesting example is provided by the unoccupied, electron
resonance states in Au adatom and addimers~\cite{NilWalPerHo} and
longer Au ad-chains~\cite{WalNilHo,NilWalHo,NilWalHo03,NazQuiHo}
assembled on a NiAl(110) surface, which were revealed by imaging
the differential conductance for various biases. This class of
resonance states was also identified in linear Cu chains on
Cu(111)~\cite{Foeletal04} so their existence is not limited to the
NiAl(110) surface.

The appearance of such resonance states in Au adatom structures
was rather surprising and raises several questions about their
physical nature. In the original
experiments~\cite{WalNilHo,NilWalHo}, the differential conductance
spectra and images of linear Au ad-chains were analyzed using the
quantum states of a simple, one-dimensional ``particle-in-box''
(PIB) model. The observed dispersion of the resonance states in
the Au chain was found to be free-electron like with an effective
mass being about half of the bare mass and the observed dependence
of the low energy peak position on the ad-chain length followed
closely the prediction of the PIB model. However, the origin and
the character of these resonance states were not clarified in this
model and the physical meaning of the one-dimensional ``box'' is
unclear in view of the discrete atomic structure of the ad-chains.
Thus there is a need to gain a more detailed understanding of
these resonance states from calculations and modelling of their
electronic structure.

The unoccupied, resonance states of the Au adatom structures are
excited states that are probed by adding an extra electron. In
principle, they are not directly described by a ground state
theory such as density functional theory. However, unoccupied,
Kohn-Sham states provide a useful, zero-order approximation of
such excited states~\cite{Gor99} as demonstrated by density
functional calculations of a single Au adatom and addimers with
various interatomic separations. These calculations showed that
the observed resonance structures in the STS were reproduced in
the calculated local density of states (LDOS) at the tip
apex~\cite{NilWalPerHo}. These resonances were formed in an energy
region with a depletion of LDOS and had a mixed $s$ and $p$
character.

A first attempt to understand the nature of the resonance states
of the Au ad-chains was the recent density functional study of
isolated, finite Au chains by Mills and coworkers~\cite{Miletal}.
They argued that the observed resonance states derived from
unoccupied states with $\pi$ character of the isolated chains.
This argument resolved the issue that the observed lowest energy
state with no nodes along the chain is unoccupied in contrast to
the states of the isolated chains with $\sigma$ character.
Furthermore, they introduced a three-dimensional
``particle-in-a-cylinder'' model to rationalize all STM
observations. However, this study did not address the effects of
the interaction of the states in the Au chains with the substrate
states.

In this paper, we present a density functional study of the
resonance states of some linear Au adatom structures adsorbed
along two inequivalent directions on the NiAl(110) surface. The
focus is primarily on the infinite Au ad-chains of which one has
the same orientation along the $[001]$ direction as an assembled
chain~\cite{WalNilHo,NilWalHo}, whereas the other infinite, Au
ad-chain is oriented along the $[1\bar{1}0]$ direction. The
substantially larger interatomic distance of the latter ad-chain
than the former ad-chain illustrates the effects of interatomic
interactions on the electronic states. We present also results for
a single adatom, two different ad-dimers and an ad-trimer. The
nature of the resonance states in the infinite ad-chains is
revealed by studying their evolution from the states of the
infinite, isolated Au chains with decreasing chain-surface
distance. We also discuss the calculated resonance states and the
tunnelling through the resonances states in terms of a
tight-binding, resonance model. This model clarifies the physical
meaning of the ``particle-in-box'' model in the analysis of the
resonance states.

The paper is organized in the following manner. In
Section~\ref{sec:compmeth}, we present some details of the density
functional calculations of the adatom structures and their geometric
and electronic structure are presented in Section~\ref{sec:res}. The
nature of the resonance states in the Au ad-chains and their relation
to the states of the isolated Au chains are discussed in
Section~\ref{sec:nature}. A simple, tight-binding resonance model for
the states in the adatom structures is described in the Appendix and
discussed in relation to calculated resonance energies and images in
Section~\ref{sec:TBRmodel}. Our results are compared with experimental
data in Section~\ref{sec:exp}, where we also clarify the physical
meaning of the ``particle-in-box'' model in terms of the
tight-binding, resonance model. Finally, we give some concluding
remarks in Section~\ref{sec:con}.

\section{Computational method \label{sec:compmeth}}

We have considered several linear Au adatom structures on the
NiAl(110) surface including single adatoms, two different addimers
and infinite, adchains along along either the $[001]$ direction or
the $[1\bar{1}0]$ direction and an adtrimer along the $[001]$
direction. The $[001]$ and the $[1\bar{1}0]$ directions correspond
to the $\bar{\Gamma}-\bar{Y}$ and the $\bar{\Gamma}-\bar{X}$
directions in the surface Brilliouin zone (SBZ),
respectively~\cite{HanGotPetetal} so that the corresponding chains
aligned along these two directions are henceforth referred to as
the $Y$ and $X$ chain, respectively. The density functional
calculations of the single adatoms and the two addimers have
already been briefly described in Ref.~\onlinecite{NilWalPerHo}.

The electronic and the geometric structure of the Au chains and
the trimer on NiAl(110) were studied using density functional
calculations that were carried out using the projector augmented
wave method as implemented in the {\tt VASP}
code~\cite{KreHaf,KreFur,KreJou}. The exchange and correlation
effects were represented by the generalized gradient
approximation~\cite{Peretal92} The two infinite, (periodic) chain
structures were represented by an Au adatom on a NiAl slab in two
different super cell geometries. To minimize the interactions
between the chain and its periodic images, we used an 1x4 and 6x1
surface unit cells for the $Y$ and $X$ chains, respectively. The
results for the electronic states of the $Y$ chain were found to
be well-converged for 9 substrate layers and this number of layers
was also used for the $X$ chain. The vacuum region of the super
cell was 6 layers. For the adtrimer we used a super cell with a
5x2 surface unit cell and 8 layers of NiAl. The kinetic energy
cut-off for the plane wave basis set was about 200 eV and the
surface Brillouin zone were sampled by 80, 28, and 24 $k$ points
for the $Y$, $X$ ad-chains, and the adtrimer, respectively. The Au
atom and the two outermost layers of Ni and Al atoms were fully
relaxed until the residual forces were less than 0.05eV/{\AA}.

To make contact with scanning tunnelling spectra, we have
calculated the local density of states (LDOS) outside the surface.
This approach is based on the Tersoff-Hamann (TH) approximation
for tunnelling in an STM junction~\cite{TerHam83}. In this
approximation, the differential conductance, $\frac{dI}{dV}(V)$,
as defined from the tunnelling current $I$ as a function of sample
bias $V$,  is given by,
\begin{eqnarray}
\frac{dI}{dV}(V) & \propto & \rho(\vec{r}_0,\epsilon = eV) \\ & = &
                 \sum_{\mu} |\psi_{\mu}(\vec{r}_0)|^2 \delta(\epsilon
                 - \epsilon_{\mu})
\label{eq:dIdVRes}
\end{eqnarray}
Here the local density of states (LDOS), $\rho(\vec{r}_0,\epsilon
= eV)$, at the position $\vec{r}_0$ of the tip apex is expanded in
wave functions $\psi_{\mu}(\vec{r}_0)$ of Kohn-Sham states $\mu$
with energy $\epsilon_\mu$ of the sample in the absence of the
tip. The continuum of one-electron states is mimicked in the slab
calculations by a Gaussian broadening of the delta function in
Eq.~\ref{eq:dIdVRes}. In principle, the range of applicability for
TH approximation to the differential conductance is limited to
small $V$. At larger $V$, one has in principle to take into
account both the change of the electronic states by the electric
field from the tip and the $V$ dependence of the tunnelling
barrier. As suggested by Lang~\cite{Lang86}, we have accounted for
the latter effect by extending the wave functions into the vacuum
region using a decay parameter set by the $V$-dependent vacuum
barrier half-way from the surface to the tip apex.

\section{Results \label{sec:res}}

In this section we will present our results for the LDOS of the
various adatom structures in the progression of adatoms,
ad-dimers, linear ad-trimers to infinite chains of Au atoms on the
NiAl(110) surface. The effects of the adatom-adatom distance on
the resonance states are revealed by considering the alignment of
the addimer and the ad-chain along the two inequivalent $X$ and
$Y$ directions of the NiAl(110) surface as defined in
Section~\ref{sec:compmeth}.

We begin by presenting the results for the geometries of the
various linear adatom structures studied in this work. Following
the suggestion by the STM experiments, we started our structural
optimization by placing the Au adatoms in the short Ni bridge
sites. In the optimization for the ad-dimers and the ad-trimer, we
find that the adatom-adatom interactions are weak compared to the
adatom-substrate interactions so that all adatoms are located
laterally within 0.03 {\AA} from the short Ni bridge positions.
Thus the adatom-adatom distance is about 2.9 {\AA} and 4.1 {\AA}
for the linear, adatom structures along the $Y$ and the $X$
directions, respectively. However, the structures with the short
adatom-adatom distance tend to increase slightly the
adatom-surface distance and relax slightly some surface Ni atom
positions. The adatom-surface distance increases from 1.95 {\AA}
for the single adatom to 2.05 {\AA} for the ad-chain along the $Y$
direction. The bare NiAl(110) surface was found to be rumpled with
the Al rows being displaced 0.18 {\AA} farther out from the
surface than the Ni rows in good agreement with earlier
experimental and theoretical studies~\cite{DavNoo,HanGotPetetal}.
The only substantial adatom-induced substrate relaxation was found
for the linear structures along the $Y$ direction and involved a
downward relaxation by about 0.1 {\AA} of the Ni atoms that are
coordinated to two Au adatoms.

\begin{figure}
\includegraphics[width=7cm]{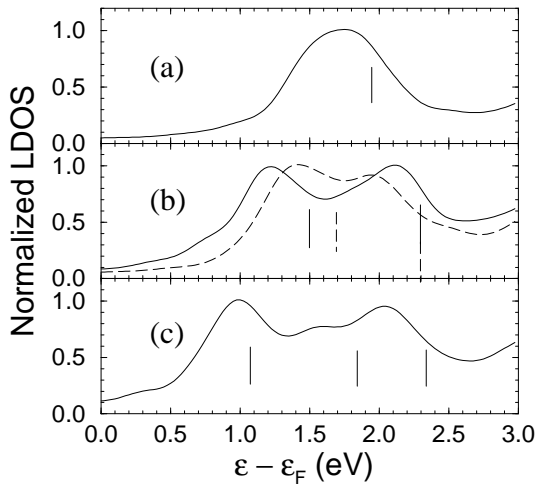}
\caption{Local density of states (LDOS) of (a) a single Au adatom
and (b) Au ad-dimers on NiAl(110), and (c) Au ad-trimer. (b) Solid
and dashed lines is the LDOS for the ad-dimer along the $Y$ and
$X$ directions, respectively. The LDOS corresponds to a
tip-surface distance of about 7 {\AA} and in a position either (a)
on top of the adatom, (b) laterally displaced 1 {\AA} away from an
adatom along the dimer axis, or (c) on top of an edge adatom. The
measured peak energies as taken from
Refs.~\protect\onlinecite{NilWalPerHo,NazQuiHo} are also indicated
by vertical bars} \label{fig:LDOS0123adatoms}
\end{figure}

The results for the LDOS of the bare NiAl(110) surface, the single
adatom and the two ad-dimers are shown in
Figs.~\ref{fig:LDOS0123adatoms}. As already reported and discussed
in Ref.~\onlinecite{NilWalPerHo}, the LDOS of the single Au adatom
exhibits a single resonance peak in the unoccupied LDOS at about
1.71 eV, which is split into a resonance doublet in the LDOS for
the two ad-dimers. The result for the LDOS of the clean surface
showed a depletion in the structure-less LDOS in this energy range
and exhibited a sharp onset around 2.5 eV above the Fermi energy.
The physical origin of these resonance states will be discussed
further in Section~\ref{sec:nature}. From the character of the
LDOS images at the peak energies, the low-lying state in the
resonance doublet of the ad-dimer was shown to be a symmetric
combination and an anti-symmetric combination of the resonance
states of the single adatoms in accordance with a simple,
two-state model.  The larger resonance splitting for the ad-dimer
along the $Y$ direction than along the $X$ direction is simply
caused by the stronger adatom-adatom interaction for the shorter
ad-dimer than for the longer ad-dimer.

\begin{figure}
\includegraphics[width=6cm]{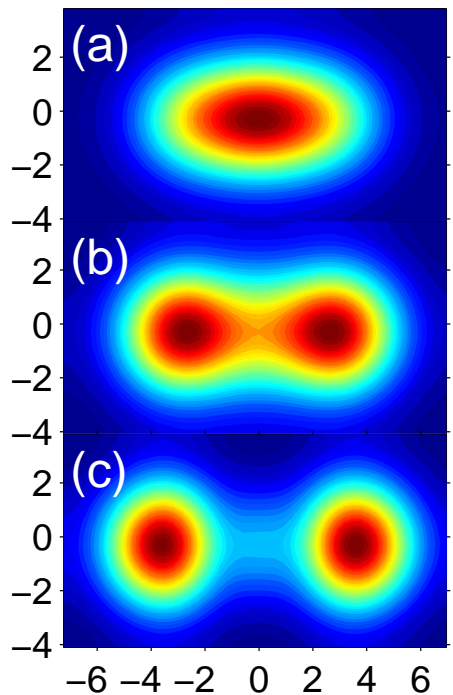}
\caption{Local density of states images of resonance states for
the Au trimer on NiAl(110).  (a) First state at 1.0 eV, (b) second
state at 1.6 eV, and (c) third state at 2.0 eV . The origin is
located at the center atom of the ad-trimer. Same tip-surface
distance as in Fig.~\protect\ref{fig:LDOS0123adatoms}}
\label{fig:LDOSTrimerImages}
\end{figure}

In the case of the linear ad-trimer, the interactions among the
three adatom resonances give rise to a resonance triplet, as shown
in Fig.~\ref{fig:LDOS0123adatoms}(c). The maximum resonance energy
splitting is about 1.0 eV and is larger than the energy splitting
of 0.9 eV for the $Y$ ad-dimer. The characters of the resonance
states are also revealed by LDOS images at the resonance peak
energies, as shown in Fig.~\ref{fig:LDOSTrimerImages}. Note that
the image for the lowest lying resonance state has no nodal
planes, whereas the other states has an increasing number of nodal
planes with increasing resonance energy. The resonance energy
positions and characters of these states and the strong intensity
of the end lobes in the LDOS image of the highest lying resonance
state will be discussed in Section~\ref{sec:TBRmodel}.

\begin{figure}
\includegraphics[width=6cm, angle = -90]{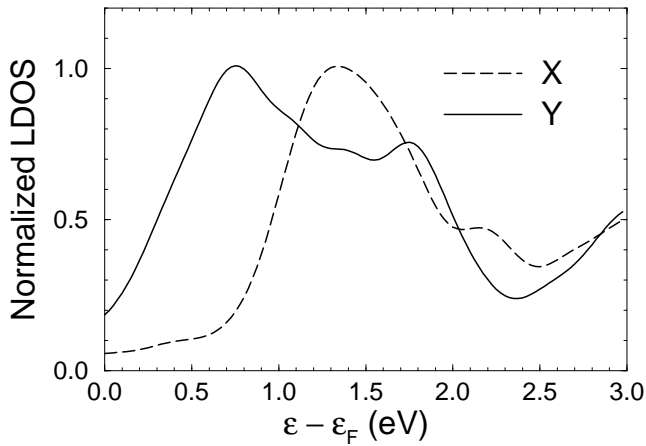}
\caption{Local density of states of the Au chains on NiAl(110)
along the (a) $Y$ and (b) $X$ directions. The tip is above an Au
atom at the same distance as in
Fig.~\protect\ref{fig:LDOS0123adatoms}.} \label{fig:LDOSChains}
\end{figure}

\begin{figure}
\includegraphics[width=8cm]{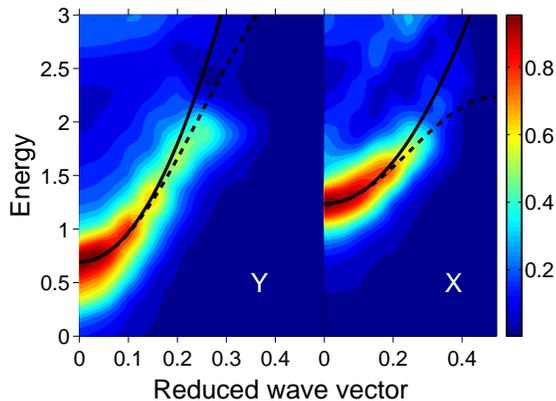}
\caption{Wave-vector--resolved local density of states of the Au
chains on NiAl(110) along the (a) $Y$ and (b) $X$ directions. The
dashed lines are dispersions as obtained from a nearest
neighboring tight binding model, whereas the solid lines are the
corresponding free particle dispersions with the same effective
mass.} \label{fig:LDOSWaveVectorChains}
\end{figure}

In the case of longer ad-chains, the interactions among the
increasing number of adatom resonance states should eventually
give rise to a band of resonance states for the infinite
ad-chains. The formation of such a band is shown in the LDOS in
Fig.~\ref{fig:LDOSChains} for the infinite ad-chains along the $Y$
and $X$ directions.  The stronger adatom-adatom interactions for
the $Y$ ad-chain than for the $X$ ad-chain results in a larger
bandwidth width for the $Y$ ad-chain than for the $X$ ad-chain.
The dispersions of the bands of resonance states are revealed by
the resolution of the LDOS over wave vectors in the
one-dimensional Brilluion zones (LBZ) of the two ad-chains. In
Fig.~\ref{fig:LDOSWaveVectorChains}, we show contour plots of
these wave-vector resolved LDOS. These results show that the LDOS
exhibit a resonance structure that disperses with the wave vector
up to the middle of LBZ where the dispersion levels off and the
resonance becomes ill-defined. For small wave vectors the
dispersion is free-electron like with an effective mass of 0.65
and 0.91 $m_e$ for the $Y$ and $X$ ad-chains, respectively.

\section{Discussion}

In this section we will discuss the physical origin of the resonance
states in the Au adatom structures on NiAl(110). This discussion
concerns primarily the formation of the resonance states in the
one-dimensional Au ad-chains. The energy positions and the LDOS images
of the resonance states of the adatom structures will be discussed in
a simple, tight-binding, resonance model.  This model enables us also
to discuss the LDOS images of ad-chains with arbitrary lengths.

\subsection{Nature of the resonance states in the Au ad-chains
\label{sec:nature}}

The existence of resonance states in Au adatom structures on the
NiAl(110) surfaces raises several questions concerning the
relation of these states to the states of the isolated adatom
structures and the role played by the electronic states of the
substrate. These questions are addressed here by scrutinizing the
resonance states of the infinite ad-chains. We begin by discussing
the states of the isolated Au chains.

\begin{figure}
\includegraphics[width=6cm, angle = -90]{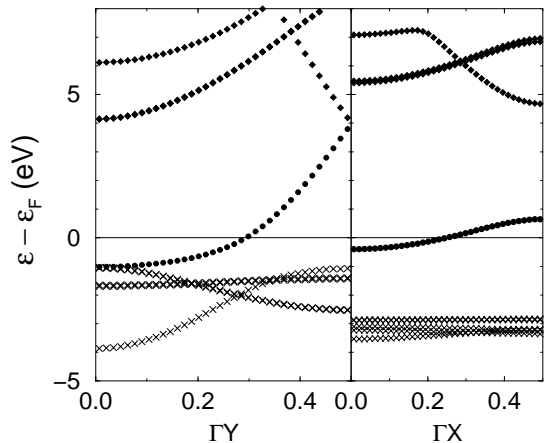}
\caption{Band structure of the isolated Au chains with same
interatomic distances as the ad-chains along the (a) $Y$ and (b)
$X$ directions. Bands with predominantly $s$, $p$, and $d$
character are marked by solid circles, diamonds and crosses,
respectively.} \label{fig:BandChains}
\end{figure}

In Figs.~\ref{fig:BandChains}(a) and (b), we show the calculated
band structures for the isolated Au chains with same geometry as
the ad-chains along the $Y$ and the $X$ directions. These results
are easily understood in terms of the electronic structure of an
isolated Au atom as obtained from spin-unpolarized density
functional calculations. The five-fold, spatially degenerate 5$d$
states of an Au atom are fully occupied with an energy 1.8 eV
below the energy of the singly occupied 6$s$ state. The
three-fold, spatially degenerate 6$p$ states are unoccupied and
located about 5 eV above the 6$s$ state. When forming an infinite,
chain these atomic states overlap and form bands. In the case of
the $X$ chain, the interatomic distance is about 4.1 {\AA} and the
atomic states form narrow and non-overlapping bands with
predominantly atomic character as shown in
Fig.~\ref{fig:BandChains}(b). Only the 6$p$ states show an
appreciable band width because they are close in energy to the
vacuum level so that the corresponding wave functions are extended
and have a large overlap. For the $Y$ chain with a shorter
interatomic distance of about 2.9 {\AA}, the dispersion of the
5$d$ and 6$s$ states increase substantially compared to the $X$
chain and develop a mixed character. The behavior of these band
structures does not provide a simple answer to the question which
atomic states are involved in the formation of the unoccupied
resonance states in the ad-chain. The 6$p$ bands are unoccupied
but are far away from the Fermi level and the 6$s$ band is
half-occupied.

\begin{figure}
\includegraphics[width=6cm, angle = -90]{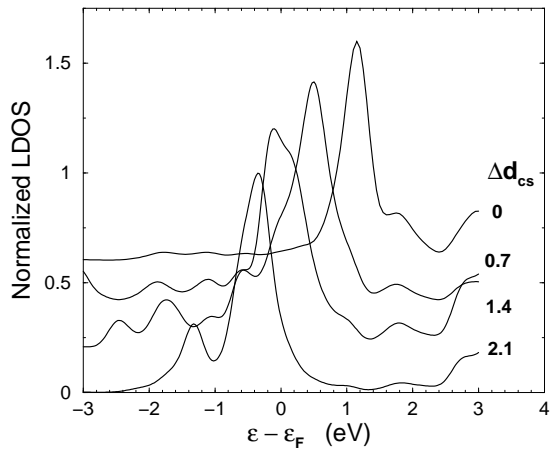}
\caption{Evolution of local density of states of the infinite, Au
chain on NiAl(110) along the $Y$ direction with outward
displacement $\Delta d_{cs}$ of the ad-chain from its equilibrium
distance. The tip--ad-chain distance is kept at a fixed distance
of 3 {\AA} above an Au atom.} \label{fig:LDOSChainXDist}
\end{figure}

To understand how the resonance states of the Au ad-chains are
related to the states of the isolated Au chains, we have
calculated the evolution of the LDOS and the partial DOS for an Au
atom in the $Y$ ad-chain at the $\Gamma$ point of the LBZ as a
function of the outward rigid displacement $\Delta d_{cs}$ of the
chain from its equilibrium position. In
Fig.~\ref{fig:LDOSChainXDist}, we show the calculated LDOS at a
fixed distance of 3 {\AA} outside the chain as a function of
$\Delta d_{cs}$. At the largest $\Delta d_{cs}$ of 2.1 {\AA}, the
chain-substrate coupling is weak and the LDOS have a prominent
peak that is close in energy to the occupied states of $s$
character of the isolated chain. For decreasing $d_{cs}$, this
state of dominant $s$ character of the chain shifts upward in
energy and broadens into the unoccupied resonance state at the
equilibrium position of the chain. Thus there is no indication in
these results for the LDOS that this resonance state evolve from a
6$p_z$ state of the Au chain that is broaden into a resonance and
shifted down continuously in energy with decreasing $\Delta
d_{cs}$. However, the result that the occupied 6$s$ state of the
chain at the $\Gamma$ point turns into a narrow unoccupied
resonance upon adsorption appears to be indicate that the ad-chain
is not charge neutral. This conflicting result is resolved by
scrutinizing the evolution of the partial DOS.

\begin{figure}
\includegraphics[width=6cm]{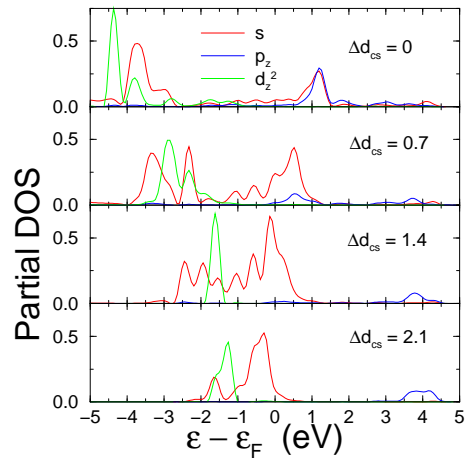}
\caption{Partial wave decomposition of the density of states for
an Au atom in the chain on NiAl(110) along the $Y$ direction for
different outward displacements, $\Delta d_{cs}$, of the ad-chain
from its equilibrium ad-chain--surface distance, as indicated in
the figure.} \label{fig:PartDOSChainXDist}
\end{figure}

In Fig.~\ref{fig:PartDOSChainXDist}, we show the evolution of the
partial DOS for an Au atom with $s$, $p_z$, and $d_{z^2}$
characters at the $\Gamma$ point in the chain BZ as a function of
$\Delta d_{cs}$. At the largest $\Delta d_{cs}$ of 2.1{\AA}, the
chain is essentially decoupled from the substrate and the partial
DOS exhibit peaks at energies close to the energies of the states
of the isolated Au chain. The state at about 0.5 eV below the
Fermi level has $s$ character and broadens appreciably with
decreasing $\Delta d_{cs}$ and develops both a resonance structure
in the unoccupied states and resonance structures in the occupied
states well below the Fermi level. Thus the formation of both
occupied and unoccupied states shows that the formation of the
resonance with $s$ character in the unoccupied states is not in
conflict of the ad-chain being neutral. The partial DOS shows also
that the resonance state has a strong $p_z$ character at the
equilibrium position but no significant $d_{z^2}$ character. The
development of the $p_z$ character of the resonance state with
decreasing $\Delta d_{cs}$ is not correlated with the 6$p_z$ state
of the isolated, chain but involves rather a polarization. This
latter state broadens substantially into a wide band upon
adsorption with decreasing $\Delta d_{cs}$ and does not shift
continuously down in energy.

\begin{figure}
\includegraphics[width=6cm]{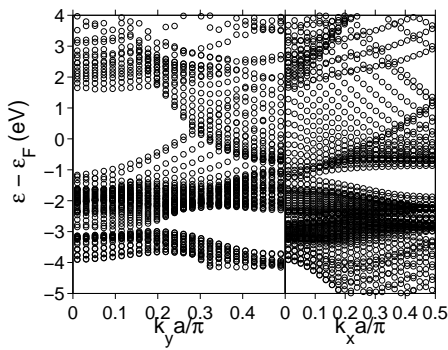}
\caption{Projected band structure of the isolated NiAl(110)
surface as a function of the reduced wave vector along the $Y$ and
$X$ directions. Only states with even character with respect to
these directions has been displayed. The bands were calculated for
an 18 layer slab.} \label{fig:BandSurface}
\end{figure}

The calculated band structure of the bare NiAl(110) surface shows
that there are projected band gaps in a energy region where the
resonance state of the ad-chain X at the $\Gamma$ point is formed.
In Fig.~\ref{fig:BandSurface}, we show the calculated electronic
states of a slab with 18 substrate layers along the $\Gamma-X$ and
$\Gamma-Y$ directions in the surface BZ (SBZ) of NiAl(110). Only
states that are even with respect to the symmetry planes spanned
by these directions in the SBZ and the surface normal are shown in
Fig.~\ref{fig:BandSurface}. The $\Gamma$ point in the LBZ of the
ad-chain $X$ corresponds to the line of $k$ points along the
$\Gamma-X$ direction in the SBZ. Along this line there is a
projected band gap in an energy region that embraces the resonance
energy about half way to the LBZ boundary which show that this
resonance state is indeed a resonance since it overlap with the
bulk states.

\begin{figure}
\includegraphics[width=6cm]{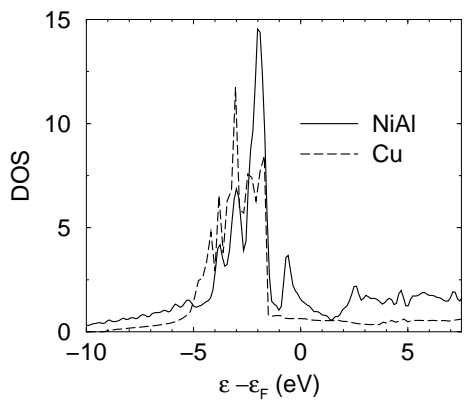}
\caption{Calculated bulk density of states of NiAl(110) (solid
line) and Cu (dashed line).} \label{fig:bulkdosNiAlCu}
\end{figure}

A depletion of density of states in the energy region of the Au
adatom-induced resonance states is revealed in the bulk DOS. In
Fig.~\ref{fig:bulkdosNiAlCu}, we compare the calculated DOS for
bulk NiAl and Cu. In the energy range of about -5 to -2 eV the DOS
is dominated by the $d$ states. In the case of NiAl, the
transition metal atoms are surrounded by a simple metal atom
resulting in a much narrower $d$ band than for Cu.  At larger
energies the DOS are dominated by states with $sp$ character. In
the case of NiAl there is a depletion in the DOS in energy range
from about 0 to 2.5 eV, whereas Cu shows no such depletion of the
DOS. Note that this depletion is not a necessary condition for the
formation of resonance state because similar states have been
shown to exist in Cu adatom structures on
Cu(111)~\cite{Foeletal04}. In the latter case the resonance state
is formed in the prominent $sp$ band gap of the Cu(111) surface.

\subsection{Simple model for resonance states and images
\label{sec:TBRmodel}}

To gain a better understanding of the interactions among the
resonances states and the LDOS images for finite chains, we have
analyzed our results in a simple tight binding, resonance (TBR)
model with an $s$-wave approximation for the tails of the
resonance wave functions in the vacuum region. This model is
defined in the Appendix. The TBR model explains the success of and
justifies the ``particle-in-the-box'' analysis of the observed
scanning tunnelling spectra for the Au chains.

We begin by investigating to what extent the TBR model can
describe the calculated resonance energies of the various adatom
structures. One obvious choice of tight-binding parameters is
based on the calculated energies of the resonance doublet of the
ad-dimers. In the case of the ad-dimer $Y$, these energies are 1.2
and 2.1 eV, which are reproduced by an on-site energy $\epsilon_0$
= 1.65 eV and a nearest-neighboring off-site energy $t_1$ = -0.45
eV. Note that in this model $\epsilon_0$ corresponds to the
resonance energy for a single adatom and $\epsilon_0$ is very
close to the calculated value of 1.7 eV for the adatom. The
predicted values for the resonance energies for the resonance
triplet of the ad-trimer are then $\epsilon_0+t_1/\sqrt{2}$ =
1.01, $\epsilon_0$=1.65, and $\epsilon_0-t_1/\sqrt{2}$ = 2.29 eV.
These values are in good agreement with the calculated energies
for the two lowest-lying states but for the highest-lying state
the calculated energy is too high with about 0.3 eV. In the case
of the ad-dimer $X$, the resonance energies suggests a on-site
resonance energy, $\epsilon_0$ = 1.70 eV that is close to its
value for the ad-dimer $Y$, whereas the off-site energy $t_1$ =
-0.30 eV is smaller than its value for the ad-dimer $Y$.

In our earlier study of the resonance doublets in ad-dimers with
various interatomic distances, we argued that there are both
direct and substrate-mediated contributions to the adatom-adatom
resonance interaction described by $t_1$. The origin of these
contributions is clarified by the result of the TBR model for
$t_1$ in Eq.~\ref{eq:RenormHopping}: $t_1 = t_1^D + t_1^S$ has a
direct contribution $t_1^D$ and a substrate-mediated contribution
$t_1^S$. $t_1^D = Zt_1^{(0)}$ is given by the direct interaction
between the 6$s$ atomic orbitals being renormalized by their
resonance strength $Z$. $t_1^{(0)}$ decays exponentially with the
interatomic distance $d$ whereas $t_1^S$ is long ranged and was
found to dominate $t_1$ for larger $d$.  $t_1^S$ can also have a
negative imaginary part which would enhance and diminish the
broadening for the symmetric and anti-symmetric resonance states,
respectively. However, this effect is not discernable in the
calculated LDOS for the ad-dimers.

In the case of the infinite ad-chains, the sets of TB parameters
derived from the ad-dimers give a poor description of the
dispersions of the resonance states for the $Y$ and $X$ ad-chains.
The predicted energies $\epsilon_0+2t_1$= 0.75 and 1.10 eV of the
lowest energy state of the TB model for the $Y$ and $X$ ad-chains,
respectively, are in agreement with the calculated energies of
0.69 and 1.23 eV for the corresponding ad-chains. However, these
parameters gives the wrong trend for the effective masses $m^*$
for the ad-chains, as defined by the near-parabolic behavior of
the dispersion for small wave vectors. The TB model with these
parameters gives $m^* = \hbar^2/(2a^2t_1) = 1.0 $ and 0.75 $m_e$
for the ad-chain $Y$ and $X$, respectively, whereas from the
calculated dispersion one obtains the reverse trend: $m^* = 0.65$
and 0.91 $m_e$ for the ad-chain $X$ and $Y$, respectively. The
breakdown of this model in describing both ad-dimers and ad-chains
is a consequence of substrate-mediated long-range interactions
among the resonance states, as was demonstrated by the combined
theoretical and experimental study in
Ref.~\onlinecite{NilWalPerHo} for the resonance states of
ad-dimers with various interatomic distances. However, as shown in
Fig.~\ref{fig:LDOSWaveVectorChains}, the near-parabolic dispersion
of the resonance states for wave vectors up to half of the LBZ is
described well by the effective TB parameters $\epsilon_0$ = 2.09
(1.83) and $t_1$ = -0.7 (-0.25) eV for the $Y$ ($X$) ad-chain.
Note that for larger wave vectors, the strengths of the resonance
states weakens and they become ill-defined.

\begin{figure}
\includegraphics[width=6cm]{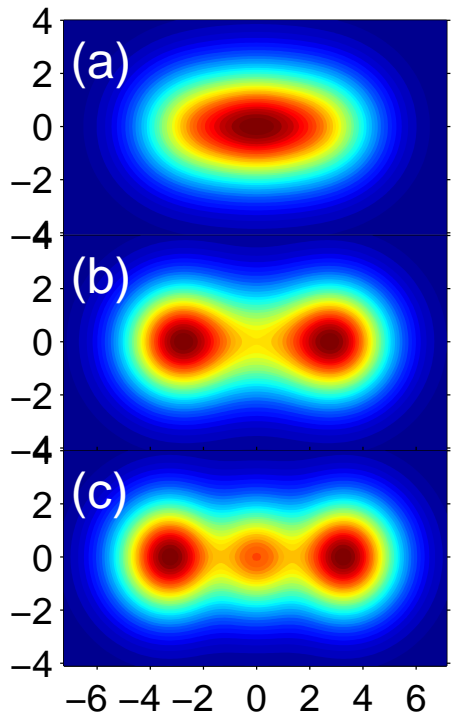}
\caption{Calculated local density of states images of a Au trimer
on NiAl(110) within the tight-binding, resonance model. First,
second and third state at (a) 1.01, (b) 1.65, and (c) 2.29 eV,
respectively. Same tip-surface distance as in
Fig.~\protect\ref{fig:LDOSTrimerImages}}
\label{fig:ModelLDOSTrimer}
\end{figure}

We have also modelled the LDOS images using the TBR model
augmented with an $s$-wave approximation for the vacuum tails of
the wave functions from each resonance state (Appendix A). In
Fig.~\ref{fig:ModelLDOSTrimer}, we show the TBR-LDOS images of the
three resonance states for the ad-trimer. The calculated images of
the two first resonance states in Fig.~\ref{fig:LDOSTrimerImages}
are rather well reproduced by this model. Note that the second
state in the TBR model is anti-symmetric and its amplitudes on the
atoms are strictly localized to the edge atoms. The absence of a
strict nodal plane in the corresponding image is caused by the
resonance broadening so there is still a contribution from the
first resonance state at the energy of the second resonance state.
At the resonance energy of the third state, the TBR-LDOS image
gives a too large contribution from the center adatom compared to
the calculated LDOS image indicating that the resonance amplitude
on the center adatom is relatively smaller to an edge adatom than
in the TBR model. Note that the LDOS images do not directly
reflect the resonance amplitudes on each adatom. For instance, in
the TBR model the only difference between the resonance amplitudes
on the adatoms between the first and the third resonance state is
the sign of the amplitude of the center adatom and its magnitude
is twice of that for an edge adatom. The large suppression of the
contribution of the center adatom to the TBR-LDOS image at the
energy of the third resonance compared to that of the edge adatom
is then caused by the destructive interference amongst the
contribution from the center adatom to the LDOS and those from the
edge atoms in contrast to a constructive interference of these
contribution to the LDOS at the energy of the first resonance
state.

\begin{figure}
\includegraphics[width=6cm, angle = -90]{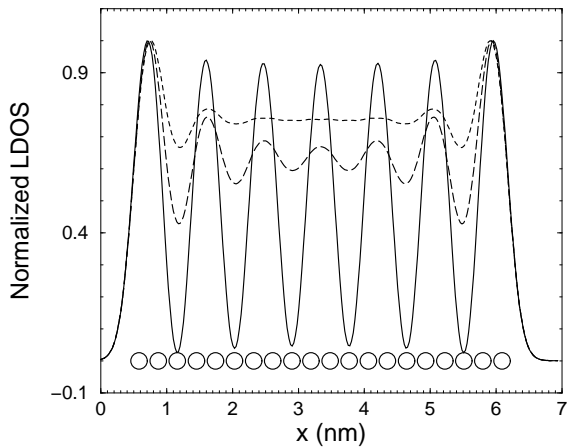}
\caption{Calculated Profiles of model local density of states
profiles of a 20 Au atom chain on NiAl(110) along $Y$ within the
tight-binding, resonance model. The profile is along the ad-chain
with the same tip-surface distance as in
Fig.~\protect\ref{fig:LDOS0123adatoms}. First, second and third
profile is at an energy of 1.39 eV corresponding to the $n=7$
resonance state with a broadening $\gamma$ equal to (a) 0.0, (b)
0.20, and (c) 0.40 eV, respectively. The circles indicate the
positions of the atoms in the ad-chain.}
\label{fig:ModelLDOSProfile20Chain}
\end{figure}

For the low-energy states of longer chains with relative long
wavelengths compared to the tip-surface distance, the resonance
broadening results in an enhancement of the LDOS at the edge
atoms. This effect is illustrated in
Fig.~\ref{fig:ModelLDOSProfile20Chain} by the results for the
TBR-LDOS profiles at an energy corresponding to the seventh
($n=7$) resonance state of a chain with 20 atoms for three
different values of the resonance broadening $\gamma$. At the
lowest value of $\gamma$ essentially only the $n=7$ state
contributes to the LDOS and there is only a minor enhancement of
the LDOS at the edge atom whereas the number of contributing
states increases with $\gamma$ and the enhancement increases at
the edge atoms. At the largest value of $\gamma$, the nodal planes
are no longer discernable.  Alternatively, this effect may be
understood simply as a reduction of the phase coherence length of
an electron propagating along the chain by the resonance
broadening.

\section{Comparison with experiments \label{sec:exp}}

The unoccupied, resonances states of chains of Au atoms of various
lengths on a NiAl(110) surface were characterized by scanning
tunnelling microscope both in an imaging and spectroscopy mode by
Nilius and coworkers~\cite{WalNilHo,NilWalHo}.  Here we are making
a comparison of the calculated resonance energies for the ad-chain
$Y$ with their experimental data. In addition, the
``particle-in-a-box'' (PIB) model used in the experimental
analysis is clarified and justified by making a direct connection
to the tight-binding resonance (TBR) model.

The calculated dispersion of the resonance states for the
infinite, ad-chain along the $Y$ direction is in agreement with
the measured dispersion from the ad-chains with eleven (Au$_{11}$)
and twenty (Au$_{20}$) adatoms. The dispersion for the Au$_{20}$
ad-chain was obtained from an analysis of the characters and the
energies of the observed resonance states in the PIB model. For
the Au$_{11}$ ad-chain the second, third and fourth resonance
state dominated the oscillations in the $\frac{dI}{dV}$ spectra at
1.3-1.4, 1.7 and 2.5 V from which the number of nodal planes and
corresponding wavelengths of these states and the box length could
be determined. A fit to a free-particle dispersion gave an
effective mass of $0.4\pm0.1 m_e$. For the Au$_{20}$ ad-chain,
several resonance states contributed to the $\frac{dI}{dV}$
spectra and the resonance energies, wavelengths and box length
were extracted from a fit of the observed STS profiles to a
superposition of several neighboring eigenstate densities of the
particle in the box. The resulting effective mass of $0.5$ m$_e$
is in close agreement with the result from the Au$_{11}$ ad-chain.
The calculated value of 0.65 m$_e$ compares favorably with these
measured values for the effective mass.

\begin{figure}
\includegraphics[width=6cm, angle = -90]{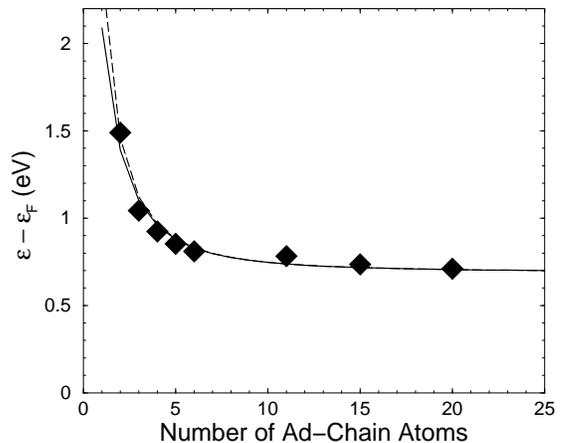}
\caption{Energy variation of the lowest energy state in the
ad-chain along the $Y$ direction as a function of number of
adatoms. The solid line is the result from the tight binding,
resonance model using the effective tight binding parameters
obtained from the calculated dispersion, whereas the dashed line
is the result using the ``particle-in-the box'' model with
parameters obtained from the tight binding resonance model. The
solid diamonds are the experimental values taken from
Refs.~\protect\onlinecite{NilWalPerHo,NazQuiHo,NilWalHo,WalNilHo}.}
\label{fig:LowEnergyState}
\end{figure}

Another more direct observation of the electronic resonance state
energies of the chain was the measurement of the energy,
$\epsilon_{min}(N)$, of the low-energy peak in the STS spectra as
a function of the number of atoms $N$ in the ad-chain.
$\epsilon_{min}(N)$ was found to follow closely the $L^{-2}$
dependence obtained in the PIB model where $L$ is the box length.
Note that the relation between $L$ and $N$ is not well-defined in
the PIB model, and the best fit was obtained using $L \approx
(N+?)a$. The calculated value of 0.69 eV for
$\epsilon_{min}(\infty)$ is very close to the band onset of 0.68
eV obtained from an extrapolation of the measured values to
$N\rightarrow\infty$. For finite ad-chains, we can make a
comparison of the calculated data with measured
$\epsilon_{min}(N)$ by using the result from the TBR model for
$\epsilon_{min}(N) = \epsilon_{min}(\infty) -
4t_1\sin^2(\frac{\pi}{2(N+1)})$. As shown in
Fig.~\ref{fig:LowEnergyState}, the agreement of the results from
this model using the effective TB parameters for the infinite,
ad-chain is in excellent agreement with the experimental data.
Note that the difference between the results of the TBR and the
free-particle limit is not significant.

This good agreement between theory and experiment for the Au
chains is somewhat fortuitous and is not granted by the use of
unoccupied Kohn-Sham states. For example, in the case of the
resonance states of the adatom, the ad-dimers and the ad-trimer
the energy difference between calculated and measured resonance
energies is typically up to 0.3 eV as shown in
Fig.~\ref{fig:LDOS0123adatoms}. As discussed in
Ref.~\onlinecite{NilWalPerHo}, a large part of this energy
difference for the resonance in the single adatom can be accounted
for by the Stark shift introduced by the electric field of the
sample bias. However, in the case of the ad-trimer, the Stark
shifts on the resonance energies are found to be small, that is,
0.014 to 0.03 eV for an external field of about 0.1V/{\AA}.

The success of the PIB model in the analysis of the observed
$\frac{dI}{dV}$ spectra and profiles, and some of its conceptual
problems are clarified by the TBR model. For example, in the PIB
model the physical meaning of the one-dimensional box and wave
functions are not clear. For wave-vectors well within the LBZ
boundaries, the TBR model gives a free-particle-like dispersion of
the PIB model. As shown in the Appendix, the envelopes of the
resonance state amplitudes on the adatoms in the TBR model are the
same as the wave function in the PIB model with $L=(N+1)a$.
Furthermore, as shown in Fig.~\ref{fig:ModelLDOSProfile20Chain},
the profile of the LDOS of a single, low-energy resonance state
along the Au$_{20}$ ad-chain in the TBR model is close to the
density of the corresponding state in the PIB model. Note that for
resonance states with higher energies and shorter wave lengths,
the LDOS densities is enhanced at the end points.

\section{Concluding remarks \label{sec:con}}

We have carried out a density functional study of the unoccupied,
resonance states in linear Au atom structures on a NiAl(110)
surface. The primary objective has been to understand the
character and the origin of these states, which were revealed and
studied by scanning tunnelling microscopy and spectroscopy (STS).
The prime focus of this study has been infinite, chains of Au
atoms along the two different orientations, $[001]$ and
$[1\bar{1}0]$, on the NiAl(110) surface. The substantial
difference in interatomic distances between these two different
orientations of the ad-chains illustrates the effects of
interatomic interactions. From a study of the evolution of the
resonance states in an ad-chain from the states of the isolated
chain, we find that the resonance states derives from the 6$s$
states of the Au adatoms, which hybridize strongly with the
substrate states and develop a substantial $p$-like polarization
perpendicular to the surface.

The calculated resonance states and their local density of states
images were analyzed in a tight-binding, resonance (TBR) model.
This analysis provides physical insight about these states. The
origin behind and the distinction between direct and
substrate-mediated interactions between the resonance states are
clarified in the TBR model. The LDOS at the end atoms of finite
ad-chains is shown to be enhanced by two different physical
mechanisms. Finally, the physical meaning of "particle-in-box"
model used in the analysis of STS for Au ad-chains is revealed in
the TBR model.

The calculated effective mass of free-particle-like dispersion of
resonance states for the infinite, Au ad-chain along the $[001]$
direction is in good agreement with the measured effective mass,
as obtained from STS of resonance states in the ad-chain with 11
and 20 Au atoms. The calculated energy of the band bottom is in
good agreement with the measured energy as obtained from an
extrapolation of the measured energies of finite ad-chains to an
infinite ad-chain length. The principal limitations of describing
the unoccupied resonance states in the Au adatom structures by
unoccupied Kohn-Sham states are revealed by the energy differences
of about 0.1- 0.3 eV between calculated and measured resonance
energies for the ad-dimers and the ad-trimer.

\begin{acknowledgments}
Financial support from the Swedish Research Council (VR) and the
Swedish Strategic Foundation (SSF) through the materials
consortium ATOMICS, and allocation of computer resources through
SNAC are gratefully acknowledged. The author are indebted to 
N. Nilius, M. Wallis, W. Ho, D. L. Mill, and R. B. Muniz for many
stimulating discussions and additional insights.
\end{acknowledgments}

\appendix*
\section{}

The proposed tight-binding, resonance model of the resonance
states in the linear adatom structures is based on a multi-site
Newns-Anderson (NA) model~\cite{And61,New69}. The Hamiltonian for
this model is given by,
\begin{widetext}
\begin{equation}
H = \sum_i \epsilon_0^{(0)}\hat{c}_i^\dagger\hat{c}_i +
    \sum_{i,j}(t_{ij}^{(0)}\hat{c}_i^\dagger\hat{c}_j + \rm{h.c.}) +
    \sum_i\sum_\mu(V_{i\mu}\hat{c}_i^\dagger\hat{c}_\mu + \rm{h.c.}) +
    \sum_\mu\epsilon_\mu\hat{c}_\mu^\dagger\hat{c}_\mu
\label{eq:MultiNAHam}
\end{equation}
\end{widetext}
Here $\hat{c}_i^\dagger$ creates an electron in an adatom state at
site $i$ of the adatom structure with energy $\epsilon_0^{(0)}$
and $t_{ij}^{(0)}$ is the direct interaction (hopping) term
between two adatoms at site $i$ and $j$, $V_{\mu i}=V_{i\mu}^*$ is
the interaction (hopping) term between an adatom state at site $i$
and a substrate state $\mu$ with energy $\epsilon_\mu$ and an
electron in this state is created by $\hat{c}_\mu^\dagger$. Note
other atomic states can be included among the substrate states.
The result for the adatom Green function for a single
site~\cite{New69} generalizes directly to the multi-site model as,
\begin{equation}
G_{ij}(\epsilon) = (G_i(\epsilon)^{-1}\delta_{ij} -
                    t_{ij}^{(0)} - \Sigma_{ij}(\epsilon)(1-\delta_{ij}))^{-1}
\label{eq:MultiAdGreen}
\end{equation}
where the Green function, $G_i(\epsilon)$, for a single adatom at
site $i$ is given by,
\begin{equation}
G_i(\epsilon) = ((\epsilon - \epsilon_0^{(0)}) - \Sigma_{ii}(\epsilon))^{-1}
\label{eq:SingleAdGreen}
\end{equation}
The effect of the substrate is represented by an energy-dependent and
complex self-energy as,
\begin{equation}
\Sigma_{ij}(\epsilon)  =
\sum_\mu \frac{V_{i\mu}V_{\mu j}}{\epsilon - \epsilon_\mu + i0^+} \\
\label{eq:SigmaDef}
\end{equation}

The resonance of the single Au adatom on the NiAl(110) surface at
the site $i$ is now modelled as a simple pole at $\epsilon_0 -
i\gamma$ in the $G_i(\epsilon)$ for the 6$s$ state with a pole
strength $Z$. Note that in this case the magnitude of $Z$ will be
substantially less than unity. Now we assume that for energies
around $\epsilon_0 + i\gamma$ the energy dependence of the
off-diagonal part of $\Sigma_{ij}(\epsilon)$ can be neglected so
that
\begin{equation}
G_{ij}(\epsilon) \approx Z ((\epsilon - \epsilon_0 + i\gamma) \delta_{ij} -
                            t_{ij})^{-1}
\label{eq:MultiAdGreenApprox}
\end{equation}
where $t_{ij}$ is a renormalized hopping term given by
\begin{equation}
t_{ij} = Z(t_{ij}^{(0)} + \Sigma_{ij}(\epsilon_0))
\label{eq:RenormHopping}
\end{equation}
This hopping term has a contribution $Zt_{ij}^{(0)}$ from the
direct interaction between 6$s$ atomic states renormalized by the
resonance strength and a contribution $Z\Sigma_{ij}(\epsilon_0)$
from substrate-mediated interactions.  Here, we restrict $t_{ij}$
to nearest-neighbor interactions. Using these approximations, the
multi-site adatom Green function reduces to an analytic form given
by,
\begin{equation}
G_{ij}(\epsilon)  =
Z\sum_k\frac{\psi_{ki}^*\psi_{ki}}{\epsilon - \epsilon_k + i\gamma_k}
\label{eq:GRes}
\end{equation}
where the energies $\epsilon_k$ and the amplitudes $\psi_{ki}$ are the
same as in the nearest neighboring TB model for a chain of states and
are given by,
\begin{eqnarray}
\epsilon_k & = & \epsilon_0 + 2{\rm Re}(t_1)\cos(ka), \ \\
\psi_{ki} & \propto & \sin(k|\vec{x}_i|) \ .
\label{eq:chainstates}
\end{eqnarray}
where $t_1$ is the nearest neighboring hopping term and
$\vec{x}_i$ are the positions of the $N$ adatoms in the linear
ad-chain with nearest neighboring adatom-adatom distance $a$, and
$k = \frac{n\pi}{(N+1)a}, n =1,2, ... N $. In contrast to the
standard TB model, $t_1$ can have a negative imaginary part from
the substrate-mediated interactions resulting in a $k$-dependent
broadening $\gamma_k$ given by,
\begin{equation}
\gamma_k = \gamma - 2{\rm Im}(t_1)\cos(ka) .
\label{eq:gammak}
\end{equation}
Finally, note that ${\Im}t_1 < 0$ will tend to increase the
broadening with decreasing $k$ and that this model contains the
ad-dimer and the ad-trimer as special cases, corresponding to $N$
= 2 and 3, respectively.

Using this TBR resonance model, we have also generated a model LDOS at
the tip apex $\vec{r}_0$ and an energy $\epsilon$ by a superposition
of localized wave functions as,
\begin{equation}
\rho(\vec{r}_0;\epsilon) = \sum_k
|\sum_i\psi_{ki}\phi(\vec{r}-\vec{x}_i;\epsilon)|
^2\frac{1}{\pi((\epsilon-\epsilon_k)^2+\gamma^2)}
\label{eq:LDOSmodel}
\end{equation}
Here we have neglected any $k$-dependent broadening. At the tip
apex, we use an $s$-wave approximation for the vacuum tail of
$\phi(\vec{r};\epsilon)$ given by,
\begin{equation}
\phi(\vec{r};\epsilon) \propto \exp(-\kappa r)/r ,
\label{eq:phitail}
\end{equation}
where $\kappa$ is the decay of the wave function in the vacuum.

\bibliography{../Articles/myarticles,../Articles/MP.9X.bib,../Articles/MP.0X.bib}

\end{document}